\documentclass[11pt,a4paper]{article}
\pdfoutput=1
\usepackage{jheppub}
\usepackage{amsmath,amssymb}
\usepackage{accents,bbm,colonequals,nicefrac}

\newcommand{\unit}[1]{\ensuremath{\,\mathrm{#1}}}

\newcommand{\ud}{\,\mathrm{d}}
\newcommand{\rd}{\mathrm{d}}
\newcommand{\bs}[1]{\boldsymbol{#1}}
\newcommand{\bfx}{\boldsymbol{x}}
\newcommand{\bfp}{\boldsymbol{p}}
\newcommand{\bfzero}{\boldsymbol{0}}

\newcommand{\ce}{\colonequals}

\newcommand{\bbC}{\mathbb{C}}

\newcommand{\umat}{\mathbbmss{1}}
\newcommand{\zmat}{\mathbb{O}}

\newcommand{\bra}[1]{\langle #1 \vert}

\newcommand{\ket}[1]{\vert #1 \rangle}

\newcommand{\abs}[1]{\lvert #1 \rvert}

\DeclareMathOperator{\e}{e}

\DeclareMathOperator{\re}{Re}

\makeatletter
\renewcommand*\env@matrix[1][c]{\hskip -\arraycolsep
  \let\@ifnextchar\new@ifnextchar
  \array{*\c@MaxMatrixCols #1}}
\makeatother

\title{Amplitudes for space-like separations and causality}

\author{S. P. Horvath,}
\author{D. Schritt}
\author{and D. V. Ahluwalia,}

\affiliation{Department of Physics and Astronomy, Rutherford Building, University of Canterbury, Private Bag 4800, Christchurch 8140, New Zealand}

\emailAdd{dharamvir.ahluwalia@canterbury.ac.nz}

\abstract{This paper explores whether quantum field theory allows the events of emission and absorption of a single particle to be separated by a space-like interval without violating Lorentz symmetries and causality. Although the answer is indeed affirmative, traditionally such effects have been considered negligible. We show that for sufficiently light mass eigenstates such processes can become significant over macroscopic length scales. A critical review of the historical literature reveals various shortcomings of the standard methods; specifically, one finds that they are restricted to states for which the expectation value of momentum vanishes. Furthermore, the results obtained here correct Feynman's analysis of this subject. A formalism is thus developed that allows the description of states with non-zero momentum, which is then applied to the OPERA and ICARUS neutrino-speed experiments. For OPERA we choose a mass in the nano electron-volt range and find that although our formalism predicts a non-zero detection probability for an early arrival time of $60 \unit{ns}$, the predicted event distribution is maximal on the light-cone. Consequently, our prediction does not reproduce the peak at $60 \unit{ns}$ reported by the OPERA collaboration. Turning to the ICARUS experiment, we note that while the collaboration reported an average time of flight that is consistent with the speed of light, the event data with its associated uncertainties nevertheless indicates that some of the detection events are separated from their corresponding emission events by a space-like interval. For a micro electron-volt mass range, this is in agreement with the here reported formalism. We thus raise the possibility of employing high-precision neutrino-speed experiments to determine the absolute masses of neutrino mass eigenstates. Finally, it is argued that the predicted space-like amplitudes are consistent with the emperical data on SN1987a.}

\keywords{Neutrino Physics}

\arxivnumber{1110.1162}

\begin{document} 

\maketitle

\section{Introduction}

This communication is a preliminary exploration of the possibility that massive particles may affect events at space-like separations. We begin with the observation that in the absence of quantum mechanical considerations Lorentz, symmetries unambiguously divide space-time into causally connected and causally disjoint regions. The boundary is occupied by massless particles. In classical theories, no massive particle can trigger a detector placed at a space-like separation from its source. In quantum theories, however, the non-commutativity of the position and momentum operators allows a classically-forbidden triggering of detectors. This possibility, as far as we know, first arose in the work of Pauli~\cite{Pauli:1941zz}. He was troubled that this could lead to a violation of the causal structure of space-time \cite{Pauli:1940zz,Pauli:1941zz,Feynman:1998rp}. It was not until the work of St\"uckelberg and Feynman \cite{Stueckelberg:1941rg,Stueckelberg:1941th,Feynman:1949hz} that the path for a consistent resolution was opened: for such processes the time ordering of events is not strictly respected, yet causality is preserved by the introduction of antiparticles. For the frames in which temporal ordering is reversed, what one would have seen as the absorption of a particle is instead observed as the emission of an antiparticle and vice versa. At the textbook level, this resolution is presented in the monographs by Feynman~\cite[chapter 18]{Feynman:1998rp} and Weinberg~\cite[section 2.13]{Weinberg:1972aa}. The understanding commonly conveyed in the modern literature is that the effects are too small to have directly observable consequences at macroscopic scales, see for example~\cite{Zee:2003mt}.

This conclusion is somewhat misleading because the scale inextricably depends on the mass of the particle. For example, Weinberg's analysis shows that the scale at which such an effect becomes significant for a proton is roughly $2\times 10^{-14} \unit{cm}$. Nevertheless, it seems to have been overlooked that if one goes down in mass by eighteen orders of magnitude to the nano electron-volt (neV) scale, then even in the Weinberg-like back-of-the-envelope calculation the range of the said effect immediately stretches to $2\times 10^{4} \unit{cm}$, that is, to two hundred meters. Given that the present data on neutrino oscillations places no lower bound on the absolute masses of neutrino mass eigenstates \cite{Nakamura:2010zzi}, this observation raises the possibility of novel neutrino phenomena. Indeed, our analysis suggests that neutrino-speed experiments may be employed to experimentally measure the absolute masses of neutrino mass eigenstates provided one of them is sufficiently light.

Even though this observation was inspired by the OPERA neutrino-speed experiment which has been recently called into question, the here-presented analysis remains valid, and may be of importance for a wider class of neutrino-speed experiments. One such example specifically discussed in this communication is the ICARUS experiment.

\section{Amplitudes for space-like separations and causality}
\label{sec:spacelikeAmplitudes}

\subsection{Calculation of the amplitudes}

To give a concrete form to the above discussion we must calculate the amplitudes for the classically-forbidden triggering of detectors mentioned in the opening paragraph of this communication. Rather than following the simplest example of a scalar field, we calculate these amplitudes in the context of a spin one-half Dirac field. This approach allows for the discussion of a number of conceptual questions that would otherwise remain unexposed. We thus consider the field~\cite{Weinberg:1995mt}
\begin{equation}
  \Psi_\ell(x) =  \sum_{\sigma}\int \frac{\ud^3p}{\left(2 \pi\right)^{3/2}}\, \sqrt{\frac{m}{p^0}} \left[u_\ell\left(\bfp, \sigma \right) \e^{-i p\cdot x} a\left(\bfp, \sigma \right) + v_\ell\left(\bfp, \sigma \right) \e^{i p\cdot x} b^\dagger\left(\bfp, \sigma \right) \right],
  \label{eqn:diracField}
\end{equation}
where the index $\ell$, $\ell\in\{1,2,3,4\}$, labels the four components of the well known Dirac field $\Psi(x)$.\footnote{The notational details are given in appendix~\ref{app:notationalDetails}.} Viewing equation~\eqref{eqn:diracField} as a linear combination of creation and annihilation operators, one naturally arrives at the question: what does the operation of $\Psi^{\ast}(x)$ on $ \ket{\mathrm{vac}}$ yield?\footnote{The reader is referred to the remarks following equation \eqref{eqn:anticommutators} for the definition of $*$. We consider the operation of $\Psi^{\ast}(x)$ on $ \ket{\mathrm{vac}}$ rather than that of $\Psi(x)$ on $ \ket{\mathrm{vac}}$ purely for convenience.} In the analogous case of a scalar field, Hatfield interprets the resulting object as a state localised at $x$~\cite{Hatfield:1992rz}. In the context of equation~\eqref{eqn:diracField}, such an object is in fact a four-dimensional column with components
\begin{equation}
  \ket{x}_\ell \ce \Psi_\ell^\ast(x) \ket{\mathrm{vac}}.
  \label{eqn:positionEigenstate}
\end{equation}
The dual to $ \ket{x}_\ell$ is
\begin{equation}
  {}_\ell\bra{x'} \ce \sum_{\ell'} \bra{\mathrm{vac}}\Psi_{\ell'}(x^\prime)\gamma_{\ell' \ell}^0,
  \label{eqn:braPositionEigenstate}
\end{equation}
where $\gamma^0$ is introduced as a metric in order to yield a Lorentz invariant inner product. Therefore, the unnormalised amplitude for a particle created at $x$ to trigger a detector located at $x^\prime$ reads\footnote{This interpretation appears implicitly in Hatfield's analysis of the problem~\cite{Hatfield:1992rz}.} 
\begin{align}
  \mathcal{A}(x\to x^\prime) \ce& \sum_\ell {}_\ell\langle x' \vert x \rangle_\ell \nonumber \\ \noalign{\medskip}
  =& \sum_{\ell \ell'} \bra{\mathrm{vac}}\Psi_{\ell'} (x^\prime)\gamma_{\ell' \ell}^0\Psi_\ell^\ast(x) \ket{\mathrm{vac}} \nonumber \\ \noalign{\medskip}
  =& \bra{\mathrm{vac}}\Psi^\mathrm{T} (x^\prime)\gamma^0\Psi^\ast(x) \ket{\mathrm{vac}}.
  \label{eqn:amplitudeDefinition}
\end{align}
The separation between $x$ and $x^\prime$ may be time-like, light-like, or space-like. The last of these separations shall yield the classically-forbidden triggering amplitudes. Evaluating the right hand side of equation~\eqref{eqn:amplitudeDefinition} using equation~\eqref{eqn:diracField}, 
one obtains
\begin{equation}
  \mathcal{A}(x \to x') = \int \frac{\ud^3 p}{(2 \pi)^3} \frac{2 m}{p^0} \e^{-i p \cdot (x' - x)}.
  \label{eqn:particlePropagationAmplitudeIntegral}
\end{equation}
A similar evaluation of the amplitude for an antiparticle yields 
\begin{align}
  \bar{\mathcal{A}}(x \to x') &= \bra{\mathrm{vac}}\Psi^\dagger (x^\prime)\gamma^0\Psi(x) \ket{\mathrm{vac}} \nonumber \\ \noalign{\medskip}
  &=-\mathcal{A}(x \to x').
  \label{eqn:antiParticlePropagationAmplitudeIntegral}
\end{align}
These amplitudes, modulo the minus sign in the above equation (dictated by the fermionic nature of the Dirac field), coincide with the known result for a massive scalar field.\footnote{The apparent spin-independence of $\mathcal{A}(x \to x')$ and $\bar{\mathcal{A}}(x \to x')$, apart from the indicated sign, is most likely due to the absence of any (spin-dependent) coupling to external fields.}

The evaluation of integrals such as equation~\eqref{eqn:particlePropagationAmplitudeIntegral} is far from trivial, and was first undertaken by Dirac~\cite{Dirac:1934pam}. Following his approach, one can rewrite equation~\eqref{eqn:particlePropagationAmplitudeIntegral} in the form
\begin{equation}
  \mathcal{A}(x \to x') = -i \frac{m}{2 \pi r} \frac{\partial}{\partial r} U(r, t),
  \label{eqn:amplitudeWrtUrt}
\end{equation}
where\footnote{The here defined $U(r,t)$ differs by a sign in the exponent from the expression quoted by Dirac \cite{Dirac:1934pam}; this is consistent, because Dirac was integrating a slight variation of equation~\eqref{eqn:particlePropagationAmplitudeIntegral}.}
\begin{equation}
  U(r, t) = \frac{1}{\pi i} \int_{-\infty}^\infty \e^{-i m [r \sinh(\varphi) + t \cosh(\varphi)]} \ud \varphi.
  \label{eqn:UrtIntegral}
\end{equation}
Here $\varphi$ is the rapidity parameter 
\begin{equation}
  \abs{\bfp} = m \sinh(\varphi) \quad\mathrm{and} \quad p^0 = m \cosh(\varphi),
\end{equation}
and
\begin{equation}
  r \ce \abs{\bfx^\prime - \bfx} \quad\mbox{and} \quad t \ce \abs{x'^0 - x^0}.
\end{equation} 
To proceed further, we employ an integral representation of the Hankel function of the first kind that can be identified with integral~\eqref{eqn:UrtIntegral}. By considering a Lommel's expansion \cite[page 140]{Watson:1944} of the Hankel function of the first kind and equating it to a suitable series expansion, along with an appropriate change of variables, one obtains (in the case where the order $\nu$ of the Hankel function of the first kind is set to $\nu = 0$)
\begin{equation}
  \frac{1}{\pi i} \int_{- \infty}^\infty \e^{-i m [t \cosh(\varphi) + i r \sinh(\varphi)]} \ud \varphi = H_0^{(1)}\left(m \sqrt{t^2 - r^2}\right).
  \label{eqn:Hankel1FinalNuZero}
\end{equation}

Mathematically there also exists an alternate solution to the aforementioned series expansion in terms of the Hankel function of the second kind, however, this solution proves to be unphysical at space-like separations and equivalent up to a sign at time-like separations; consequently it is not considered. Substitution of equation~\eqref{eqn:Hankel1FinalNuZero} into equation~\eqref{eqn:amplitudeWrtUrt} yields 
\begin{equation}
  \mathcal{A}(x \to x') = - \frac{i}{2 \pi}\;\frac{m^2}{\sqrt{t^2 - r^2}} H_1^{(1)}\left(m \sqrt{t^2 - r^2} \right).
  \label{eqn:amplitudeGeneralSolution}
\end{equation}
This expression is valid for all space-time separations. If $x$ and $x'$ are separated by a purely space-like interval, amplitude~\eqref{eqn:amplitudeGeneralSolution} may be rewritten in terms of a modified Bessel function of the second kind of order $\nu = 1$
\begin{equation}
  \mathcal{A}(x \to x') = -\bar{\mathcal{A}}(x \to x') = \frac{1}{\pi^2}\frac{m^2}{\sqrt{r^2 - t^2}} K_1\left(m \sqrt{r^2 - t^2} \right).
  \label{eqn:spacelikeAmplitude}
\end{equation}
The result~\eqref{eqn:amplitudeGeneralSolution} is consistent with that of Pauli~\cite{Pauli:1941zz}, and for an analogous integral in a different context, also with that of Dirac~\cite{Dirac:1934pam}. Feynman, however, obtains a result in terms of a Hankel function of the second kind and order $\nu = 1$~\cite{Feynman:1949hz,Feynman:1998rp}. This result proves to be divergent for an increasing parameter of the space-like interval~\cite{Watson:1944}. A second error occurs when he gives an incorrect asymptotic expansion for Hankel functions of the second kind \cite[page 198]{Watson:1944}. For a space-like interval, the resulting incorrect expansion matches the asymptotic behaviour of equation~\eqref{eqn:amplitudeGeneralSolution}; however, inside the light-cone it yields an oscillatory solution that contradicts equation~\eqref{eqn:amplitudeGeneralSolution}. 

Of course the true genius in these historical investigations lies in the prediction of antiparticles. We take a brief moment to revisit this argument. First note that since $t^2 - r^2$ is Lorentz invariant, the amplitudes $\mathcal{A}(x \to x')$ and $\bar{\mathcal{A}}(x \to x')$ also exhibit the same invariance. Second, consider two space-like separated events $x$ and $x^\prime$. Since the amplitudes $\mathcal{A}(x \to x')$ and $\bar{\mathcal{A}}(x \to x')$ are non-zero there exist two classes of observers, one for which $x^{\prime 0} > x^0$ and the other for which $x^0 > x^{\prime 0}$. The temporal ordering of events is thus no longer preserved. Causality is then ensured by the following interpretation: if the first set of observers sees a fermion created at $x$ and subsequently absorbed at $x^\prime$, then the second set of observers sees an antifermion created at $x^\prime$ and absorbed at $x$, and vice versa. As far as we can infer, the credit for this insight goes to Pauli, St\"uckelberg, and Feynman. The extension of these arguments to bosons runs along similar lines.

\subsection{The massless limit}

In this section we explore the behaviour of the amplitude \eqref{eqn:spacelikeAmplitude} in the massless limit. As has been shown by Weinberg, the massless limit of the Dirac field is well defined and coincides with the field operator obtained by the demand that any Hamiltonian density constructed from the field operator must transform appropriately under Lorentz transformations \cite{Weinberg:1964ev}. Consequently, we are justified in directly taking the $m \to 0$ limit of the amplitude \eqref{eqn:spacelikeAmplitude}. For a finite space-time interval $\sqrt{r^2 - t^2}$ the argument of the modified Bessel function of the second kind will be small as $m$ approaches infinitesimal values, hence we can consider the limiting form
\begin{equation}
  K_\nu(z)\big\vert_{z\to 0} \sim \frac{1}{2} (\nu - 1)! \left(\frac{1}{2} z\right)^{-\nu},
  \label{eqn:modifiedBesselFunctionSmallArgumentLimit}
\end{equation}
for $z \in \bbC$ and $\re(\nu) > 0$ \cite{Abramowitz:1964}. It thus follows that the amplitude for a classically forbidden triggering vanishes in the $m \to 0$ limit 
\begin{equation}
  \lim_{m \to 0} \mathcal{A}(x \to x') = \lim_{m \to 0} \frac{m}{\pi^2 (r^2 - t^2)} = 0.
  \label{eqn:amplitudeMassLimit}
\end{equation}
Furthermore, for a fixed space-time interval, the amplitude \eqref{eqn:spacelikeAmplitude} vanishes for both large as well as small values of $m$ with a maximal probability dictated by the order of the space-time interval.
\begin{figure}[h!]
  \begin{center}
    \includegraphics{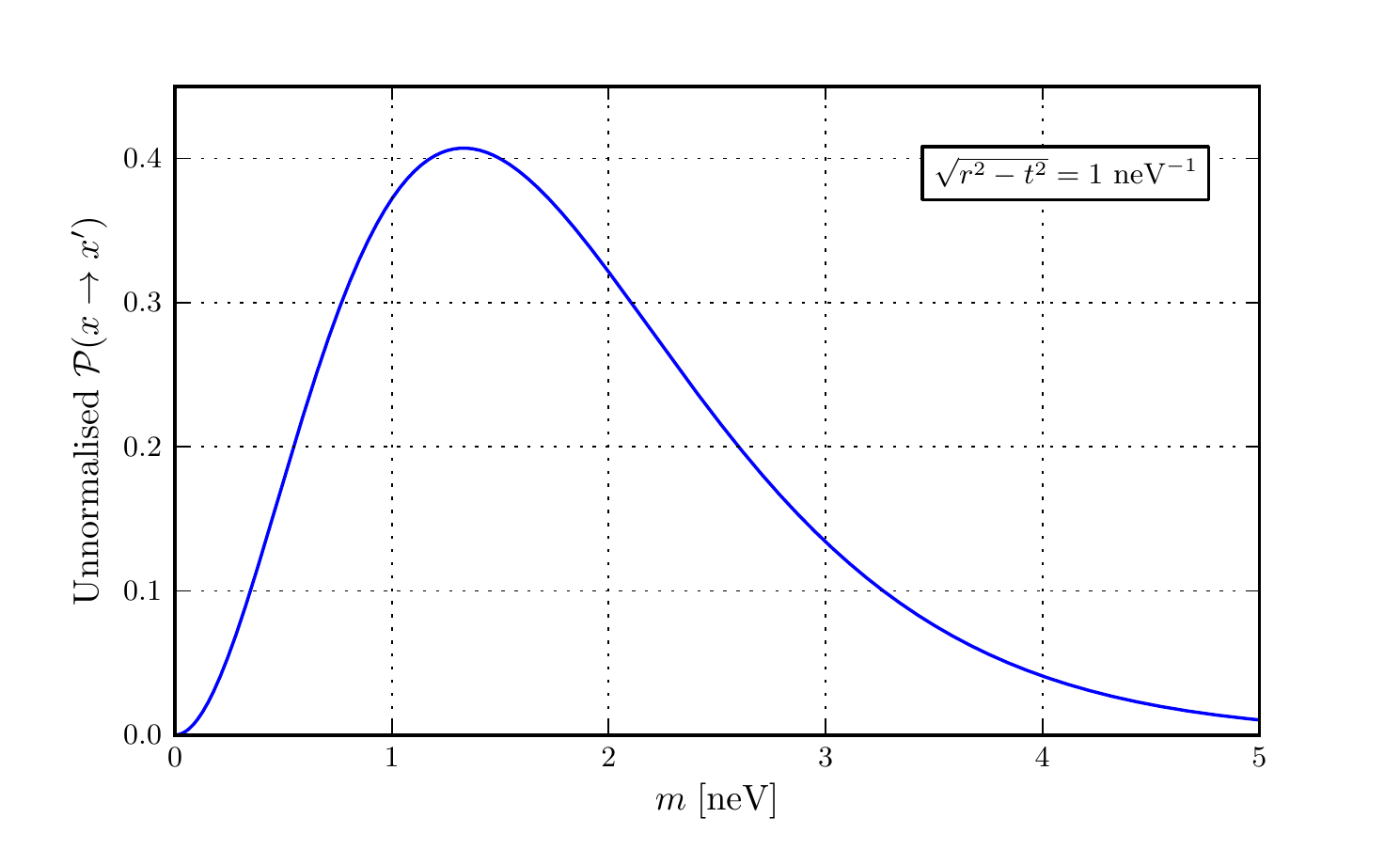}
    \caption{The plot depicts the mass dependence of the probability $\mathcal{P}(x \to x')$ for a fixed space-time interval of $\sqrt{r^2 - t^2}= 1 \unit{neV}^{-1} = 1.24 \unit{km}$. It highlights the existence of a maximal probability with respect to a variation in mass and, consequently, the smoothness of the massless limit.}
    \label{fig:stationaryPointPlot}
  \end{center}
\end{figure}
Figure \ref{fig:stationaryPointPlot} depicts the mass dependence of the unnormalised\footnote{The probability is unnormalisable due to the highly localised nature of the state that results from the application of a field operator on the vacuum (see section \ref{subsec:amplitudeInterpretation} below). The remedy is, of course, well known. One must introduce a state with a finite width in configuration space.} probability $\mathcal{P}(x \to x') :=\mathcal{A}^\ast(x \to x') \mathcal{A}(x \to x')$ for a space-time interval of $\sqrt{r^2 - t^2} = 1 \unit{neV}^{-1}$. It is possible to further qualify this result: repeating the above derivation of equation \eqref{eqn:spacelikeAmplitude} using a massless Dirac field leads to the same result as that obtained in equation \eqref{eqn:amplitudeMassLimit}.

\subsection{Interpretation of the amplitudes}
\label{subsec:amplitudeInterpretation}

In order to provide further physical interpretation for the above amplitudes a closer look at $\ket{x}$ is required. From equation \eqref{eqn:positionEigenstate} this state in component form reads 
\begin{align}
  \ket{x}_\ell &= \sum_{\sigma}\int \frac{\ud^3 p}{\left(2 \pi\right)^{3/2}}\, \sqrt{\frac{m}{p^0}}\, u^\ast_\ell\left(\bfp, \sigma \right) \e^{i p\cdot x} a^\dagger\left(\bfp, \sigma \right) \ket{\mathrm{vac}} \\ \noalign{\medskip}
  &= \sum_{\sigma}\int \frac{\ud^3p}{\left(2 \pi\right)^{3/2}}\, \sqrt{\frac{m}{p^0}}\,u^\ast_\ell\left(\bfp, \sigma \right) \e^{i p\cdot x} \ket{\bfp, \sigma}.
\end{align}
It is thus apparent that for $\ket{x}$ the expectation value of the momentum operator vanishes with infinite variance. This agrees with the expectation for a completely localised state in configuration space. Consequently, the amplitudes $\mathcal{A}(x \to x') $ and $\bar{\mathcal{A}}(x \to x')$ are only applicable to particles for which the expectation value of momentum vanishes.

The application of these amplitudes to neutrino-speed experiments \cite{adam:2011zb,Adamson:2007zzb,Antonello:2012hg} requires additional effort.\footnote{The work of Morris \cite{Morris:2011nt} considers the evolution of a wave-packet to describe a moving particle in this context. Such a description is, however, not valid in a relativistic context; see, for example, H.\ and T.\ Padmanabhan \cite[section V. A.]{Padmanabhan:2011ty} for a detailed discussion on the evolution of a relativistic wave-packet.} Before we embark on this task in section~\ref{sec:neutrinos}, we note that a graphical analysis of the probability $\mathcal{P}(x \to x')$ reveals that a particle has the highest probability of triggering a detector located at a light-like separation from its source. Furthermore, one finds that there is an asymmetry between the probability of a particle triggering a detector at a space-like separation and a time-like separation.

\section{Neutrino-speed experiments}
\label{sec:neutrinos}

We will now explore a context in which the above amplitudes have an experimentally observable effect; in particular, we consider neutrino-speed experiments.

While neutrinos produced in electroweak processes are a linear superposition of mass eigenstates, we make the simplifying assumption that the entire process corresponds to the emission and detection of a single mass eigenstate. This assumption will capture the essence of our argument. Thus, consider a single ultra-relativistic mass eigenstate produced at a time $\eta=0$ and directed towards a detector placed at a spatial distance $L$. In the absence of the discussed amplitudes, the canonical expectation is that it may be detected no sooner than $t_\mathrm{c} = L$. In the spirit of this preliminary investigation, at any given instant $\eta$ the emitted mass eigenstate may be instantaneously considered at rest (this allows the use of the amplitudes obtained above). Its classically-expected spatial location is $\eta$. One can thus ask: what is the probability amplitude that the mass eigenstate under consideration triggers the detector at a time $t_\mathrm{o} \le t_\mathrm{c}$? That is, what is the probability amplitude, $A(\Gamma)$, that the mass eigenstate triggers the detector at a time $\Gamma \ce t_\mathrm{c} - t_\mathrm{o}$ \emph{earlier} than the canonically expected time? Within the defined setting, an application of equation~\eqref{eqn:spacelikeAmplitude} leads to the result
\begin{equation}
  A(\Gamma) = \int_0^{t_\mathrm{o}} \hspace{-7pt} \rd \eta\; \frac{1}{\pi^2} \frac{m^3}{\xi(\eta)}{K_1\left[ \xi(\eta) \right]},
  \label{eqn:amplitudeWrtXi}
\end{equation}
where $\xi(\eta) := {m \sqrt{(L - \eta)^2 - (t_\mathrm{o} - \eta)^2}}$. 

In order to circumvent the unnormalisable character of ${P}(\Gamma)$ we define the ratio 
\begin{equation}
  \alpha_r(\Gamma) \ce \frac{P(\Gamma)}{P(\Gamma_{\mathrm{r}})}= \frac{A^\ast(\Gamma)A(\Gamma)}{A^\ast(\Gamma_{\mathrm{r}})A(\Gamma_{\mathrm{r}})}
  \label{eqn:probabilityRatio}
\end{equation}
where $\Gamma_{\mathrm{r}}$ is a conveniently chosen experiment-specific reference point.

\subsection{OPERA neutrino-speed result}
\label{subsec:OPERA}

\begin{figure}[hbt!]
  \begin{center}
    \includegraphics{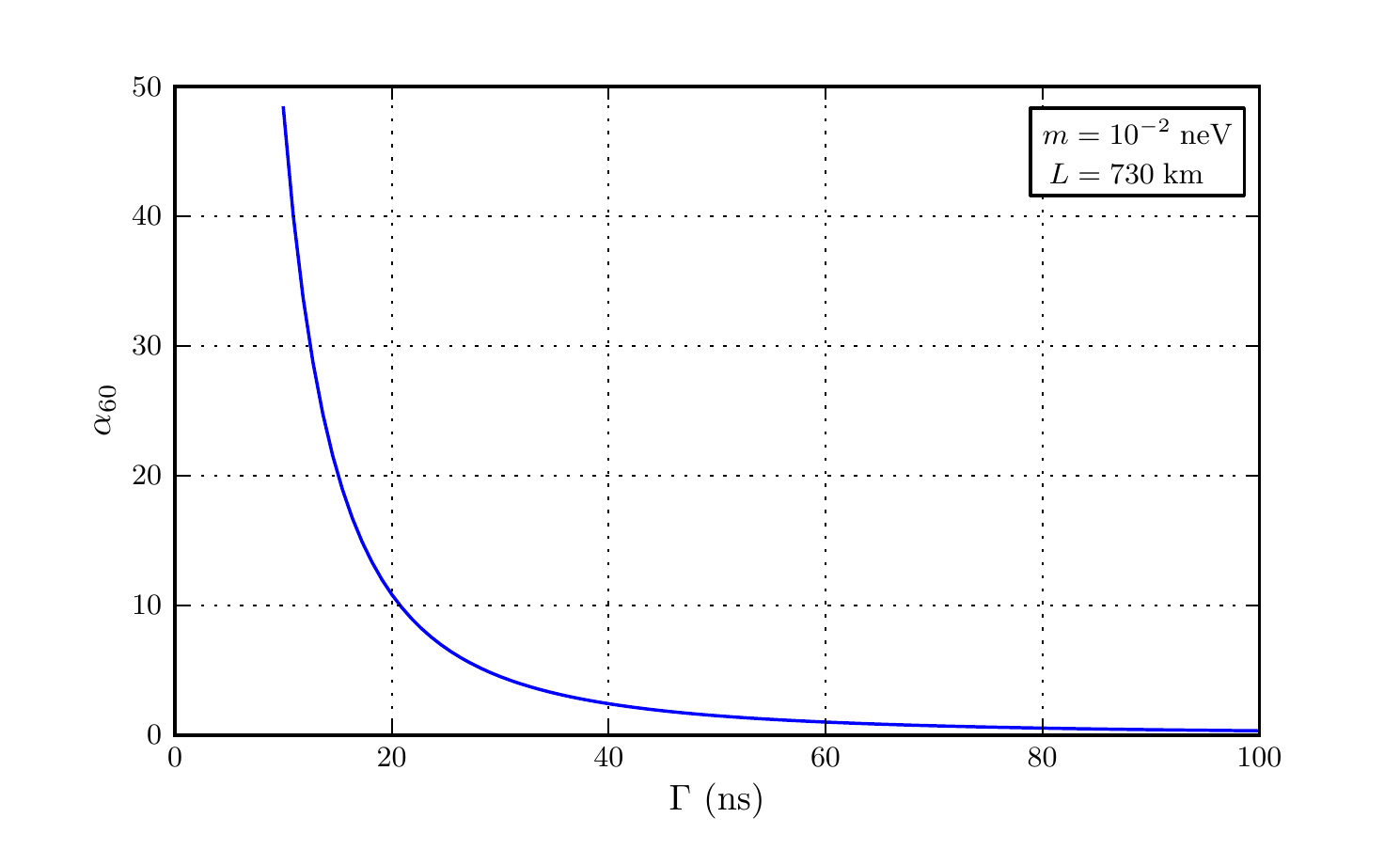}
    \caption{A graphical representation of the probability to detect a single mass eigenstate outside the light-cone. The ordinate $\alpha_{60}$ corresponds to the ratio of the probability at $\Gamma$ and the probability at $\Gamma_{60} \ce 60 \unit{ns}$.}
    \label{fig:spacelikeAmplitudeOPERA}
  \end{center}
\end{figure}

For comparison with the OPERA results, we choose $\Gamma_{60} \ce 60 \unit{ns}$ as the reference point, and use equation \eqref{eqn:probabilityRatio} to calculate $\alpha_{60}(\Gamma)$. For a fixed baseline $L = 730 \unit{km}$, as is applicable to the OPERA experiment, a detailed graphical analysis shows that as one lowers the mass in expression~\eqref{eqn:probabilityRatio} the variation of $\alpha_{60}$ with $\Gamma$ saturates for $m \lesssim 10^{-2} \unit{neV}$. This mass range is a natural choice in the context of the OPERA experiment because it maximises the space-like amplitudes in expression \eqref{eqn:amplitudeWrtXi}. A graphical representation of $\alpha_{60}(\Gamma)$ for $m = 10^{-2} \unit{neV}$ is provided in figure~\ref{fig:spacelikeAmplitudeOPERA}, which reveals that although there is a non-vanishing probability of detecting a mass eigenstate at $\Gamma\approx 60 \unit{ns}$ the predicted event distribution is larger for smaller values of $\Gamma$.\footnote{Integral \eqref{eqn:amplitudeWrtXi} was solved numerically by employing the mpmath arbitrary-precision floating-point arithmetic library \cite{mpmath}. Figures throughout this communication were produced using the matplotlib plotting tools \cite{hunter:90} in conjunction with the Numpy and Scipy scientific computation libraries \cite{numpyScipy}.}
The most expedient path to compare this event distribution with the OPERA results is to draw attention to the data pertaining to the short-bunch wide-spacing beam~\cite[figure 18]{adam:2011zb}. For the $20$ events of this type recorded, the collaboration reported 
\begin{itemize}
  \item[a.] a mean value of $\Gamma = (62.1 \pm 3.7) \unit{ns}$,
  \item[b.] that all the $20$ events occur roughly in the range $40 \unit{ns} \le \Gamma \le 90 \unit{ns}$,
  \item[c.] that there are no events in the range for $\Gamma \le 40 \unit{ns}$.
\end{itemize}
This does not match our theoretically predicted event distribution. 

Since the original announcement, the OPERA collaboration has unearthed two sources of errors~\cite{Opera:2012ne}. At the time of writing it is not clear how the proverbial dust will settle \cite{Chodos:1984cy,Widom:2011md,Chashchina:2011aj,Baccetti:2011xs,AmelinoCamelia:2011dx,Cohen:2011hx,Anacleto:2011bv,Altaie:2011fv,Alfaro:2011sp,Hannestad:2011bj,Wang:2011zk,Pas:2005rb,Contaldi:2011zm,Drago:2011ua,Alexandre:2011bu}. Nevertheless, provided that there exists a sufficiently light mass eigenstate, our analysis suggests that the space-like correlations under discussion will be detectable.

\subsection{ICARUS neutrino-speed result}
\label{subsec:ICARUS}

We now turn to the neutrino-speed experiment performed using the ICARUS detector. Although the collaboration reported an \emph{average} time of flight consistent with the speed of light, the \emph{individual} events detailed in reference \cite[table 1]{Antonello:2012hg} nevertheless indicate that some of the detection events are separated from their corresponding emission events by a space-like interval. It is the distribution of such events that we wish to compare to the here-predicted event distribution. 

\begin{figure}[hbt!]
  \begin{center}
    \includegraphics{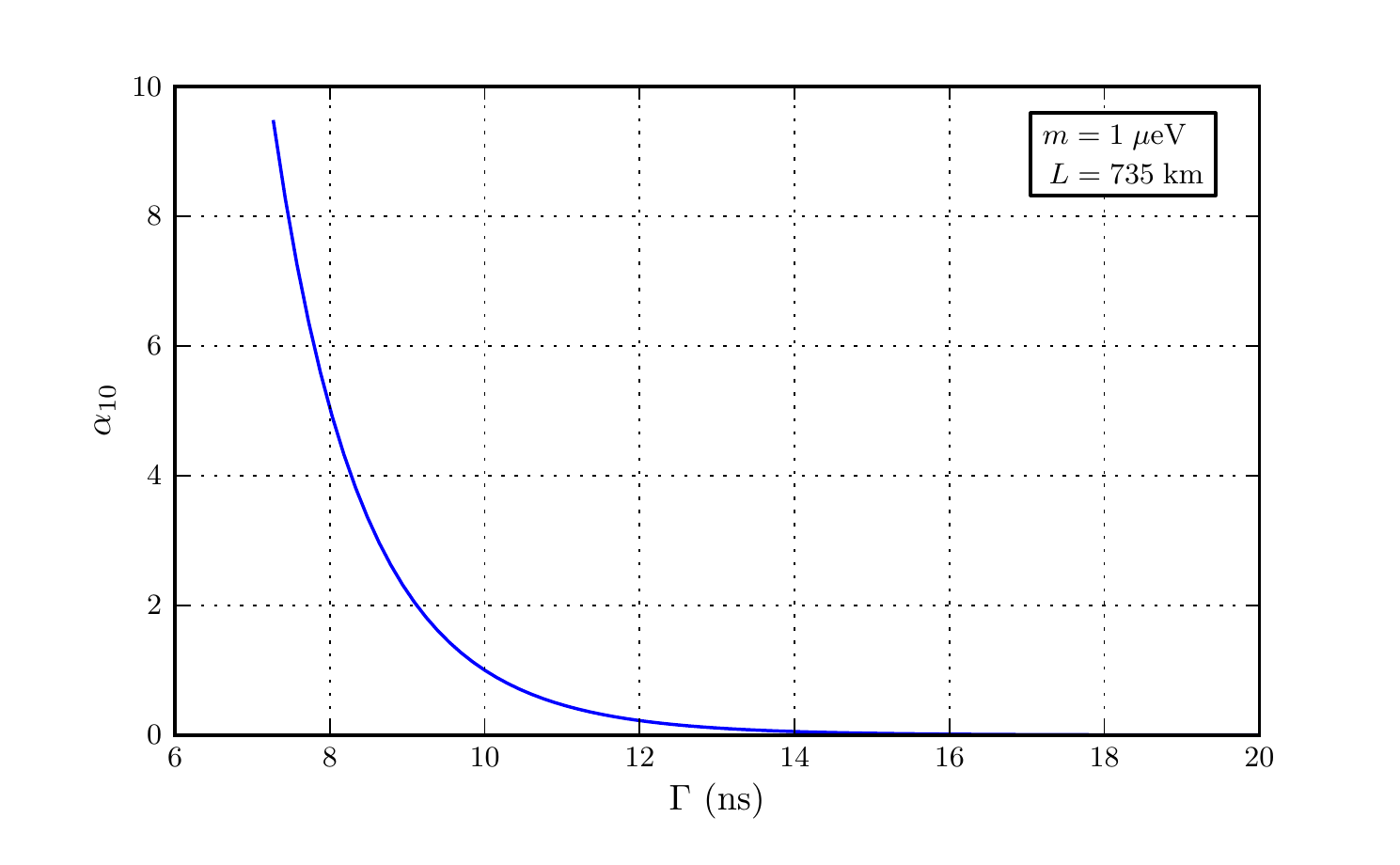}
    \caption{A graphical representation of the probability to detect a single mass eigenstate outside the light-cone. The ordinate $\alpha_{10}$ corresponds to the ratio of the probability at $\Gamma$ and the probability at $\Gamma_{10} \ce 10 \unit{ns}$.}
    \label{fig:spacelikeAmplitudeICARUS}
  \end{center}
\end{figure}

Guided by the ICARUS data, we choose 
$\Gamma_{10} \ce 10 \unit{ns}$ and compute the distribution $\alpha_{10}(\Gamma)$. To gain a qualitative insight, the results obtained for a representative mass of $m=1\; \mu\mathrm{eV}$ are depicted in figure~\ref{fig:spacelikeAmplitudeICARUS}. 
Although this choice of mass roughly reproduces the ICARUS distribution \cite[figure 3]{Antonello:2012hg}, it is not possible to perform a statistically meaningful fit given that the currently available data consists of only seven events with a subset of four events corresponding to space-like trajectories \cite{Antonello:2012hg}. 

If such event distributions were verified by employing high-precision neutrino-speed experiments, the presented formalism would provide a direct means for experimentally determining the absolute masses of neutrino mass eigenstates. We add the following caveat to this conclusion: given that the canonical flavour oscillation formalism does not take into account the mass-dependent amplitudes for space-like separations, such a determination would require a generalisation of the standard neutrino oscillation phenomenology.

\subsection{Short baseline neutrino-speed experiments} 

For short-baseline experiments, such as LSND \cite{Athanassopoulos:1996jb,Athanassopoulos:1997pv}, the amplitudes \eqref{eqn:amplitudeWrtXi} are only applicable to non-oscillation events and we will consequently restrict the discussion that follows to such events.\footnote{The LSND neutrino-oscillation data requires a mass-squared difference of the order of an $\unit{eV}^2$. It is unlikely that this particular mass-squared difference is associated with the $\unit{\mu eV}$-$\unit{neV}$ range mass eigenstates. If the latter mass eigenstates do exist, these will be detected mostly in the non-oscillations events.} For comparison with the above analysis, we consider $m = 10^{-2} \unit{neV}$ and find that a reduction of the baseline from $L^{\text{\tiny OPERA}} = 730 \unit{km}/L^{\text{\tiny ICARUS}} = 735 \unit{km}$ to $L^{\text{\tiny LSND}} = 30 \unit{m}$ yields a suppression of $\alpha_{\mathrm{r}}(\Gamma)$. Interestingly, by setting $m = 1 \unit{\mu eV}$ one finds that a reduction of the baseline to $30 \unit{m}$ now results in a practically indistinguishable $\alpha_{\mathrm{r}}(\Gamma)$. This can be understood by analysing the form of equation \eqref{eqn:amplitudeWrtXi}; the modified Bessel function of the second kind approaches zero for an increasing argument, so that the contributions to the integration of equation \eqref{eqn:amplitudeWrtXi} over large values of $\eta$ become negligible. Therefore, for a sufficiently small mass the values over which the modified Bessel function of the second kind contributes to the integration is greater than the baseline of $L^{\text{\tiny LSND}} = 30 \unit{m}$. By increasing the mass, however, all the contributing terms to the integration can be shifted to within the considered baseline.

\section{SN1987a}

Repeating the above analysis for the scenario of SN1987a by adjusting the baseline to $L^\text{\tiny SN1987a} \approx 1.8 \times 10^5\unit{ly}$~\cite{Hirata:1987hu,Bionta:1987qt,Hirata:1988ad} yields effectively indistinguishable results from those shown in figures \ref{fig:spacelikeAmplitudeOPERA} and \ref{fig:spacelikeAmplitudeICARUS}, for their respective mass scales. This may once again by understood by considering the form of equation \eqref{eqn:amplitudeWrtXi}: over large values of $\eta$, the contribution to the integration by the modified Bessel function of the second kind becomes negligible. The significance of this is that, for the considered mass scales, any early arrival time due to this effect would be of the order of nano seconds, even for baselines of astrophysical magnitude. This is consistent with the data recorded for SN1987a.

\section{Conclusion}

A preliminary quantum field theoretic investigation of the possible detection of a particle across a space-like separated interval was undertaken. We have found that for sufficiently light mass eigenstates space-like amplitudes become significant for macroscopic length scales. Furthermore, we have shown that such amplitudes are only applicable to states with a vanishing momentum expectation value. Therefore, a formalism incorporating such amplitudes for a moving mass eigenstate was developed and applied to the OPERA and ICARUS neutrino-speed experiments. This yielded a mass dependent distribution with a maximal detection probability centred around a light-like trajectory. 

For comparison with the OPERA experiment a choice of $m \sim 10^{-2} \unit{neV}$ yielded a significant detection probability for events with an early arrival time of $60 \unit{ns}$; however, given that the predicted event distribution is maximal on the light-cone it did not reproduce the peak at $60 \unit{ns}$ of the OPERA event distribution. In order to compare the developed formalism with the ICARUS experiment a choice of $m \sim 1 \unit{\mu eV}$ yielded a visual agreement between the predicted and the observed event distributions. Given that the currently available data consists of only seven data points it was, however, not possible to calculate a statistically meaningful fit. Nevertheless, this raises the possibility of employing high-precision neutrino-speed experiments to determine the absolute masses of neutrino mass eigenstates. Furthermore, it was shown that the predicted space-like amplitudes are consistent with the empirical data on SN1987a. 

\appendix

\section{Notational details}
\label{app:notationalDetails}

In equation~\eqref{eqn:diracField} $\sigma$ takes on the values $\pm 1/2$, and $m$ refers to the mass of the particles and antiparticles described by the field $\Psi(x)$; $p^0$ is the time component of the energy-momentum four vector $p^\mu$, and equals $\sqrt{\bfp^2+m^2}$. Any such $p^\mu$ can be expressed in terms of the standard four-momentum $k^\mu\ce (m,\bs{0})$ by the usual Lorentz transformation:
\begin{equation}
  p^\mu = L^\mu_{\phantom{\mu}\nu} (p)\, k^\nu.
\end{equation}
The expansion coefficients given in equation~\eqref{eqn:diracField} differ from those given in reference \cite{Weinberg:1995mt} by a factor of $\sqrt{m/p^0}$, such that 
\begin{align}
  u(\bfp, \sigma) &= D\left( L(p)\right) u(\bfzero, \sigma), \\ \noalign{\medskip}
  v(\bfp, \sigma) &= D\left( L(p)\right) v(\bfzero, \sigma),
\end{align}
with
\begin{equation}
  D\left( L(p)\right) = \sqrt{\frac{E+m}{2 m}}
  \begin{pmatrix}
    \umat +\frac{\bs{\sigma}\cdot\bfp}{E+m} & \zmat \\
    \zmat & \umat -\frac{\bs{\sigma}\cdot\bfp}{E+m}
  \end{pmatrix}.
\end{equation}
Here, $\bs{\sigma} \ce (\sigma_x, \sigma_y, \sigma_z)$, are the standard Pauli matrices; $\umat$ and $\zmat$ are the $2\times 2$ identity and null matrices, respectively. The expansion coefficients at rest are defined by 
\begin{alignat}{4}
  u(\bfzero,1/2) &= \frac{1}{\sqrt{2}} 
  \begin{bmatrix}[r]
    1 \\ 0 \\ 1 \\ 0
  \end{bmatrix}, \qquad &u(\bfzero,-1/2) &= \frac{1}{\sqrt{2}} 
  \begin{bmatrix}[r]
    0 \\ 1 \\ 0 \\ 1 
  \end{bmatrix}, \\ \noalign{\medskip}
  v(\bfzero,1/2) &= \frac{1}{\sqrt{2}} 
  \begin{bmatrix}[r]
    0 \\ 1 \\ 0 \\ -1 
  \end{bmatrix}, \qquad &v(\bfzero, -1/2) &= \frac{1}{\sqrt{2}} 
  \begin{bmatrix}[r]
    -1 \\ 0 \\ 1 \\0 
  \end{bmatrix},
  \label{eqn:expansionCoefficients}
\end{alignat}
where the index $\ell$ introduced in equation~\eqref{eqn:diracField} is the row index of the above defined expansion coefficients. In reference to equation~\eqref{eqn:braPositionEigenstate}
\begin{equation}
  \gamma^0 = \begin{pmatrix}
    \zmat & \umat \\
    \umat & \zmat
  \end{pmatrix}.
\end{equation}
Furthermore, the vacuum $\ket{\mathrm{vac}}$ is Lorentz invariant with $\langle\mathrm{vac}\vert\mathrm{vac}\rangle =1$, and the annihilation and creation operators satisfy the usual fermionic anticommutation relations
\begin{align}
  \left\{a(\bfp^\prime,\sigma^\prime), a^\dagger(\bfp,\sigma)\right\} &= \delta^3(\bfp^\prime-\bfp) \delta_{\sigma^\prime\sigma} \nonumber \\ \noalign{\medskip}
  \left\{a^\dagger(\bfp^\prime,\sigma^\prime), a^\dagger(\bfp,\sigma)\right\} &= 0 \nonumber \\ \noalign{\medskip}
  \Big\{a(\bfp^\prime,\sigma^\prime), a(\bfp,\sigma)\Big\} &= 0.
  \label{eqn:anticommutators}
\end{align}
In equations \eqref{eqn:positionEigenstate} and \eqref{eqn:amplitudeDefinition}, $\ast$ represents complex conjugation and $\mathrm{T}$ denotes matrix transposition. The following important convention is then used 
\begin{equation} 
  a^\ast(\bfp,\sigma) \equiv a^\dagger(\bfp,\sigma). 
  \label{eqn:astDaggerEquivalence}
\end{equation}
The counterpart of equations \eqref{eqn:anticommutators} and \eqref{eqn:astDaggerEquivalence} holds for $b(\bfp,\sigma)$ and $b^\dagger(\bfp,\sigma)$.

\acknowledgments

The authors are thankful to Cheng-Yang Lee, Terry Goldman, Ben Martin, Kane O'Donnell and Oscar Macias-Ramirez for discussion, the anonymous referee for a helpful and insightful referee report, and S. Elizabeth Graham for proof reading. SPH thanks Rick Beatson for a very helpful consultation on integral representations of Hankel functions.

\providecommand{\href}[2]{#2}\begingroup\raggedright\endgroup


\begin{thebibliography}{10}

\bibitem{Pauli:1941zz}
W.~Pauli, {\it {Relativistic field theories of elementary particles}},  {\em
  Rev. Mod. Phys.} {\bf 13} (1941) 203--232.

\bibitem{Pauli:1940zz}
W.~Pauli, {\it {The connection between spin and statistics}},  {\em Phys. Rev.}
  {\bf 58} (1940) 716--722.

\bibitem{Feynman:1998rp}
R.~P. Feynman, {\em {The theory of fundamental processes}}.
\newblock {Westview Press, Cambridge, Massachusetts, USA}, {1998}.

\bibitem{Stueckelberg:1941rg}
E.~St{\"u}ckelberg, {\it {Remarks on the creation of pairs of particles in the
  theory of relativity}},  {\em Helv. Phys. Acta} {\bf 14} (1941) 588--594.

\bibitem{Stueckelberg:1941th}
E.~St{\"u}ckelberg, {\it {La mecanique du point materiel en theorie de
  relativite et en theorie des quanta}},  {\em Helv. Phys. Acta} {\bf 15}
  (1942) 23--37.

\bibitem{Feynman:1949hz}
R.~Feynman, {\it {The theory of positrons}},  {\em Phys. Rev.} {\bf 76} (1949)
  749--759.

\bibitem{Weinberg:1972aa}
S.~Weinberg, {\em Gravitation and cosmology: principles and applications of the
  general theory of relativity}.
\newblock John Wiley \& Sons, New York, USA, 1972.

\bibitem{Zee:2003mt}
A.~Zee, {\em {Quantum field theory in a nutshell}}.
\newblock Princeton University Press, Princeton, New Jersey, USA, 2003.

\bibitem{Nakamura:2010zzi}
{\bf Particle Data Group} Collaboration, K.~Nakamura {\em et.~al.}, {\it
  {Review of particle physics}},  {\em J. Phys. G} {\bf G37} (2010) 075021.

\bibitem{Weinberg:1995mt}
S.~Weinberg, {\em {The quantum theory of fields. Vol. 1: foundations}}.
\newblock {Cambridge University Press, Cambridge, UK}, 1995.

\bibitem{Hatfield:1992rz}
B.~Hatfield, {\em {Quantum field theory of point particles and strings}}.
\newblock {Westview Press, Cambridge, Massachusetts, USA}, 1992.

\bibitem{Dirac:1934pam}
P.~A.~M. Dirac, {\it {Discussion of the infinite distribution of electrons in
  the theory of the positrons}},  {\em Proc. Camb. Phil. Soc.} {\bf 30} (1934)
  150--163.

\bibitem{Watson:1944}
G.~H. Watson, {\em Theory of Bessel functions}.
\newblock Cambridge University Press, Cambridge, UK, 1944.

\bibitem{Weinberg:1964ev}
S.~Weinberg, {\it {Feynman rules for any spin. 2. Massless particles}},  {\em
  Phys. Rev.} {\bf 134} (1964) B882--B896.

\bibitem{Abramowitz:1964}
M.~Abramowitz and I.~A. Stegun, {\em Handbook of mathematical functions}.
\newblock Dover Publications Inc., New York, USA, 1964.

\bibitem{adam:2011zb}
T.~Adam {\em et.~al.}, {\it {Measurement of the neutrino velocity with the
  OPERA detector in the CNGS beam}},
  \href{http://xxx.lanl.gov/abs/1109.4897}{{\tt arXiv:1109.4897}}.

\bibitem{Adamson:2007zzb}
{\bf MINOS} Collaboration, P.~Adamson {\em et.~al.}, {\it {Measurement of
  neutrino velocity with the MINOS detectors and NuMI neutrino beam}},  {\em
  Phys. Rev.} {\bf D76} (2007) 072005,
  [\href{http://xxx.lanl.gov/abs/0706.0437}{{\tt arXiv:0706.0437}}]. 6 pages, 3
  figures, to be submitted to PRD. Added discussion and expanded text.

\bibitem{Antonello:2012hg}
{\bf ICARUS} Collaboration, M.~Antonello {\em et.~al.}, {\it {Measurement of
  the neutrino velocity with the ICARUS detector at the CNGS beam}},
  \href{http://xxx.lanl.gov/abs/1203.3433}{{\tt arXiv:1203.3433}}.

\bibitem{Morris:2011nt}
T.~R. Morris, {\it {Off-shell OPERA neutrinos}},
  \href{http://xxx.lanl.gov/abs/1110.3266}{{\tt arXiv:1110.3266}}.

\bibitem{Padmanabhan:2011ty}
H.~Padmanabhan and T.~Padmanabhan, {\it {Non-relativistic limit of quantum
  field theory in inertial and non-inertial frames and the principle of
  equivalence}},  {\em Phys. Rev.} {\bf D84} (2011) 085018,
  [\href{http://xxx.lanl.gov/abs/1110.1314}{{\tt arXiv:1110.1314}}].

\bibitem{mpmath}
F.~Johansson {\em et.~al.}, {\em mpmath: a {P}ython library for
  arbitrary-precision floating-point arithmetic (version 0.14)}, February,
  2010.
\newblock {\tt http://code.google.com/p/mpmath/}.

\bibitem{hunter:90}
J.~D. Hunter, {\it Matplotlib: a 2d graphics environment},  {\em Computing in
  Science \& Engineering} {\bf 9} (2007) 90--95.

\bibitem{numpyScipy}
T.~E. Oliphant, {\it Python for scientific computing},  {\em Computing in
  Science \& Engineering} {\bf 9} (2007) 10--20.

\bibitem{Opera:2012ne}
E.~S. Reich, {\it {Timing glitches dog neutrino claim}},  {\em Nature} {\bf
  483} (2012) 17.

\bibitem{Chodos:1984cy}
A.~Chodos, A.~I. Hauser, and V.~Kostelecky, {\it {The neutrino as a tachyon}},
  {\em Phys. Lett.} {\bf B150} (1985) 431.

\bibitem{Widom:2011md}
A.~Widom, J.~Swain, and Y.~Srivastava, {\it {Space-like motions of quantum zero
  mass neutrinos}},  \href{http://xxx.lanl.gov/abs/1111.7181}{{\tt
  arXiv:1111.7181}}.

\bibitem{Chashchina:2011aj}
O.~Chashchina and Z.~Silagadze, {\it {Breaking the light speed barrier}},
  \href{http://xxx.lanl.gov/abs/1112.4714}{{\tt arXiv:1112.4714}}.

\bibitem{Baccetti:2011xs}
V.~Baccetti, K.~Tate, and M.~Visser, {\it {Inertial frames without the
  relativity principle}},  \href{http://xxx.lanl.gov/abs/1112.1466}{{\tt
  arXiv:1112.1466}}.

\bibitem{AmelinoCamelia:2011dx}
G.~Amelino-Camelia, G.~Gubitosi, N.~Loret, F.~Mercati, G.~Rosati, {\em
  et.~al.}, {\it {OPERA-reassessing data on the energy dependence of the speed
  of neutrinos}},  \href{http://xxx.lanl.gov/abs/1109.5172}{{\tt
  arXiv:1109.5172}}.

\bibitem{Cohen:2011hx}
A.~G. Cohen and S.~L. Glashow, {\it {Pair creation constrains superluminal
  neutrino propagation}},  {\em Phys. Rev. Lett.} {\bf 107} (2011) 181803,
  [\href{http://xxx.lanl.gov/abs/1109.6562}{{\tt arXiv:1109.6562}}].

\bibitem{Anacleto:2011bv}
M.~Anacleto, F.~Brito, and E.~Passos, {\it {Supersonic velocities in
  noncommutative acoustic black holes}},
  \href{http://xxx.lanl.gov/abs/1109.6298}{{\tt arXiv:1109.6298}}.

\bibitem{Altaie:2011fv}
M.~Altaie, {\it {Does the superluminal neutrino uncover torsion?}},
  \href{http://xxx.lanl.gov/abs/1111.0286}{{\tt arXiv:1111.0286}}.

\bibitem{Alfaro:2011sp}
J.~Alfaro, {\it {Superluminal neutrinos and the standard model}},
  \href{http://xxx.lanl.gov/abs/1110.3540}{{\tt arXiv:1110.3540}}.

\bibitem{Hannestad:2011bj}
S.~Hannestad and M.~S. Sloth, {\it {Apparent faster than light propagation from
  light sterile neutrinos}},  \href{http://xxx.lanl.gov/abs/1109.6282}{{\tt
  arXiv:1109.6282}}.

\bibitem{Wang:2011zk}
P.~Wang, H.~Wu, and H.~Yang, {\it {Superluminal neutrinos and monopoles}},
  \href{http://xxx.lanl.gov/abs/1110.0449}{{\tt arXiv:1110.0449}}.

\bibitem{Pas:2005rb}
H.~Pas, S.~Pakvasa, and T.~J. Weiler, {\it {Sterile-active neutrino
  oscillations and shortcuts in the extra dimension}},  {\em Phys. Rev.} {\bf
  D72} (2005) 095017, [\href{http://xxx.lanl.gov/abs/hep-ph/0504096}{{\tt
  hep-ph/0504096}}].

\bibitem{Contaldi:2011zm}
C.~R. Contaldi, {\it {The OPERA neutrino velocity result and the
  synchronisation of clocks}},  \href{http://xxx.lanl.gov/abs/1109.6160}{{\tt
  arXiv:1109.6160}}.

\bibitem{Drago:2011ua}
A.~Drago, I.~Masina, G.~Pagliara, and R.~Tripiccione, {\it {The hypothesis of
  superluminal neutrinos: comparing OPERA with other data}},
  \href{http://xxx.lanl.gov/abs/1109.5917}{{\tt arXiv:1109.5917}}.

\bibitem{Alexandre:2011bu}
J.~Alexandre, J.~Ellis, and N.~E. Mavromatos, {\it {On the possibility of
  superluminal neutrino propagation}},
  \href{http://xxx.lanl.gov/abs/1109.6296}{{\tt arXiv:1109.6296}}.

\bibitem{Athanassopoulos:1996jb}
{\bf LSND} Collaboration, C.~Athanassopoulos {\em et.~al.}, {\it {Evidence for
  $\bar\nu_\mu\to\bar\nu_e$ oscillations from the LSND experiment at LAMPF}},
  {\em Phys. Rev. Lett.} {\bf 77} (1996) 3082--3085,
  [\href{http://xxx.lanl.gov/abs/nucl-ex/9605003}{{\tt nucl-ex/9605003}}].

\bibitem{Athanassopoulos:1997pv}
{\bf LSND} Collaboration, C.~Athanassopoulos {\em et.~al.}, {\it {Evidence for
  $\nu_\mu\to\nu_e$ oscillations from LSND}},  {\em Phys. Rev. Lett.} {\bf 81}
  (1998) 1774--1777, [\href{http://xxx.lanl.gov/abs/nucl-ex/9709006}{{\tt
  nucl-ex/9709006}}].

\bibitem{Hirata:1987hu}
{\bf KAMIOKANDE-II} Collaboration, K.~Hirata {\em et.~al.}, {\it {Observation
  of a neutrino burst from the supernova SN 1987a}},  {\em Phys. Rev. Lett.}
  {\bf 58} (1987) 1490--1493.

\bibitem{Bionta:1987qt}
R.~Bionta, G.~Blewitt, C.~Bratton, D.~Casper, A.~Ciocio, {\em et.~al.}, {\it
  {Observation of a neutrino burst in coincidence with supernova SN 1987a in
  the large Magellanic cloud}},  {\em Phys. Rev. Lett.} {\bf 58} (1987) 1494.

\bibitem{Hirata:1988ad}
K.~Hirata, T.~Kajita, M.~Koshiba, M.~Nakahata, Y.~Oyama, {\em et.~al.}, {\it
  {Observation in the Kamiokande-II detector of the neutrino burst from
  supernova SN 1987a}},  {\em Phys. Rev.} {\bf D38} (1988) 448--458.

\end{thebibliography}
\end{document}